\documentclass[twocolumn]{aastex631}

\usepackage[utf8]{inputenc}
\usepackage{mathtools}
\usepackage{booktabs}
\usepackage{multirow}

\usepackage{amsmath}
\usepackage{pifont}
\newcommand{\co}{\ifmmode {\mathrm{CO}} \else CO\fi}
\newcommand{\coeighteen}{\ifmmode {\mathrm{C}^{18}\mathrm{O}} \else C$^{18}$O\fi}
\newcommand{\thirteenco}{\ifmmode {^{13}\mathrm{CO}} \else $^{13}$CO\fi}
\newcommand{\hdens}{\ifmmode {n_{\mathrm{H}_{2}}} \else $n_{\mathrm{H}_{2}}$\fi}
\newcommand{\hdensLog}{\ifmmode {n_{0,\mathrm{H}_{2}}} \else $n_{0,\mathrm{H}_{2}}$\fi}
\newcommand{\tkin}{\ifmmode {\mathrm{T}_{\mathrm{kin}}} \else $\mathrm{T}_{\mathrm{kin}}$\fi}
\newcommand{\coldens}{\ifmmode{N_{\mathrm{CO}}} \else $N_{\mathrm{CO}}$\fi}
\newcommand{\width}{\ifmmode{\sigma} \else $\sigma$ \fi}
\newcommand{\xthirteen}{\ifmmode{{X}_{12/13}} \else ${X}_{12/13}$\fi}
\newcommand{\xeighteen}{\ifmmode{{X}_{13/18}} \else ${X}_{13/18}$\fi}
\newcommand{\hydrogen}{\ifmmode {\mathrm{H}_{2}} \else H$_{2}$\fi}
\newcommand{\bff}{\ifmmode {\Phi} \else $\Phi$\fi}
\newcommand{\conv}{\ifmmode {\textrm{CO-to-H}_{2}}   \else CO-to-H$_2$ \fi}
\newcommand{\alphaco}{\ifmmode {\alpha_{\mathrm{CO}}}   \else $\alpha_{\mathrm{CO}}$ \fi}
\usepackage{amsmath}
\usepackage{color,colortbl}
\usepackage{makecell}
\begin{document}

\newcommand{\Princeton}{\affiliation{Department of Astrophysical Sciences, Princeton University, 4 Ivy Lane, Princeton, NJ 08544, USA}}

\newcommand{\McMaster}{\affiliation{Department of Physics and Astronomy, McMaster University, 1280 Main Street West, Hamilton, ON L8S 4M1, Canada}}

\newcommand{\CITA}{\affiliation{Canadian Institute for Theoretical Astrophysics (CITA), University of Toronto, 60 St George Street, Toronto, ON M5S 3H8, Canada}}

\newcommand{\OSU}{\affiliation{Department of Astronomy, The Ohio State University, 140 West 18th Avenue, Columbus, OH 43210, USA}}

\newcommand{\CCAPP}{\affiliation{Center for Cosmology and Astroparticle Physics (CCAPP), 191 West Woodruff Avenue, Columbus, OH 43210, USA}}

\newcommand{\Alberta}{\affiliation{Department of Physics, University of Alberta, Edmonton, AB T6G 2E1, Canada}}

\newcommand{\Cologne}{\affiliation{I. Physikalisches Institut der Universität zu Köln, Zülpicher Straße 77, 50937, Köln, Germany}} 

\newcommand{\ANU}{\affiliation{Research School of Astronomy and Astrophysics, Australian National University, Canberra, ACT 2611, Australia}}

\newcommand{\Arizona}{\affiliation{Steward Observatory, University of Arizona, 933 North Cherry Avenue, Tucson, AZ 85721, USA}}

\newcommand{\ASIAA}{\affiliation{Institute of Astronomy and Astrophysics, Academia Sinica, No. 1, Sec. 4, Roosevelt Road, Taipei 106216, Taiwan}}

\newcommand{\ASTROThreeD}{\affiliation{ARC Centre of Excellence for All Sky Astrophysics in 3 Dimensions (ASTRO 3D), Australia}}

\newcommand{\Bonn}{\affiliation{Argelander-Institut f\"ur Astronomie, Universit\"at Bonn, Auf dem H\"ugel 71, 53121 Bonn, Germany}}

\newcommand{\Carnegie}{\affiliation{Observatories of the Carnegie Institution for Science, 813 Santa Barbara Street, Pasadena, CA 91101, USA}}

\newcommand{\CCA}{\affiliation{Center for Computational Astrophysics, Flatiron Institute, 162 Fifth Avenue, New York, NY 10010, USA}}

\newcommand{\CfA}{\affiliation{Center for Astrophysics $\mid$ Harvard \& Smithsonian, 60 Garden Street, Cambridge, MA 02138, USA}}

\newcommand{\CITEVA}{\affiliation{Centro de Astronomía (CITEVA), Universidad de Antofagasta, Avenida Angamos 601, Antofagasta, Chile}}

\newcommand{\CNRS}{\affiliation{CNRS, IRAP, 9 Av. du Colonel Roche, BP 44346, F-31028 Toulouse cedex 4, France}}

\newcommand{\Connecticut}{\affiliation{Department of Physics, University of Connecticut, 196A Auditorium Road, Storrs, CT 06269, USA}}

\newcommand{\COOL}{\affiliation{Cosmic Origins Of Life (COOL) Research DAO, coolresearch.io}}

\newcommand{\EPFL}{\affiliation{Institute of Physics, Laboratory for galaxy evolution and spectral modelling, EPFL, Observatoire de Sauverny, Chemin Pegais 51, 1290 Versoix, Switzerland}}

\newcommand{\ESO}{\affiliation{European Southern Observatory, Karl-Schwarzschild Stra{\ss}e 2, D-85748 Garching bei M\"{u}nchen, Germany}}

\newcommand{\Gent}{\affiliation{Sterrenkundig Observatorium, Universiteit Gent, Krijgslaan 281 S9, B-9000 Gent, Belgium}}

\newcommand{\Hawaii}{\affiliation{Institute for Astronomy, University of Hawaii, 2680 Woodlawn Drive, Honolulu, HI 96822, USA}}

\newcommand{\Heidelberg}{\affiliation{Astronomisches Rechen-Institut, Zentrum f\"{u}r Astronomie der Universit\"{a}t Heidelberg, M\"{o}nchhofstra\ss e 12-14, D-69120 Heidelberg, Germany}}

\newcommand{\IAC}{\affiliation{Instituto de Astrof\'isica de Canarias, C/ V\'ia L\'actea s/n, E-38205, La Laguna, Spain}}

\newcommand{\ICRAR}{\affiliation{International Centre for Radio Astronomy Research, University of Western Australia, 35 Stirling Highway, Crawley, WA 6009, Australia}}

\newcommand{\INAF}{\affiliation{INAF -- Osservatorio Astrofisico di Arcetri, Largo E. Fermi 5, I-50157, Firenze, Italy}}

\newcommand{\IPAC}{\affiliation{Caltech-IPAC, 1200 E. California Blvd. Pasadena, CA 91125, USA}}

\newcommand{\IPARC}{\affiliation{Instituto de F\'{\i}sica de Part\'{\i}culas y del Cosmos IPARCOS, Facultad de Ciencias F\'{\i}sicas, Universidad Complutense de Madrid, E-28040, Spain}}

\newcommand{\IRAM}{\affiliation{Institut de Radioastronomie Millim\'etrique (IRAM), 300 Rue de la Piscine, F-38406 Saint Martin d'H\`eres, France}}

\newcommand{\ITA}{\affiliation{Universit\"{a}t Heidelberg, Zentrum f\"{u}r Astronomie, Institut f\"{u}r Theoretische Astrophysik, Albert-Ueberle-Str 2, D-69120 Heidelberg, Germany}}

\newcommand{\IWR}{\affiliation{Universit\"{a}t Heidelberg, Interdisziplin\"{a}res Zentrum f\"{u}r Wissenschaftliches Rechnen, Im Neuenheimer Feld 205, D-69120 Heidelberg, Germany}}

\newcommand{\JHU}{\affiliation{Department of Physics and Astronomy, The Johns Hopkins University, Baltimore, MD 21218, USA}}

\newcommand{\Kansas}{\affiliation{Department of Physics and Astronomy, University of Kansas, 1251 Wescoe Hall Drive, Lawrence, KS 66045, USA}}

\newcommand{\LAM}{\affiliation{Aix Marseille Univ, CNRS, CNES, LAM (Laboratoire d’Astrophysique de Marseille), Marseille, France}}

\newcommand{\Leiden}{\affiliation{Leiden Observatory, Leiden University, P.O. Box 9513, 2300 RA Leiden, The Netherlands}}

\newcommand{\Liverpool}{\affiliation{Astrophysics Research Institute, Liverpool John Moores University, IC2, Liverpool Science Park, 146 Brownlow Hill, Liverpool L3 5RF, UK}}

\newcommand{\Lyon}{\affiliation{Univ Lyon, Univ Lyon 1, ENS de Lyon, CNRS, Centre de Recherche Astrophysique de Lyon UMR5574, F-69230 Saint-Genis-Laval, France}}

\newcommand{\Maryland}{\affiliation{Department of Astronomy and Joint Space-Science Institute, University of Maryland, 4296 Stadium Drive, College Park, MD 20742, USA}}
\newcommand{\UMarry}{\affiliation{Department of Astronomy, University of Maryland, 4296 Stadium Drive, College Park, MD 20742, USA}}

\newcommand{\Yebes}{\affiliation{Centro de Desarrollos Tecnológicos, Observatorio de Yebes (IGN), 19141 Yebes, Guadalajara, Spain}}
\newcommand{\MPE}{\affiliation{Max-Planck-Institut f\"{u}r extraterrestrische Physik, Giessenbachstra{\ss}e 1, D-85748 Garching, Germany}}

\newcommand{\MPIA}{\affiliation{Max-Planck-Institut f\"{u}r Astronomie, K\"{o}nigstuhl 17, D-69117, Heidelberg, Germany}}

\newcommand{\Nagoya}{\affiliation{Department of Physics, Nagoya University, Furo-cho, Chikusa-ku, Nagoya, Aichi 464-8602, Japan}}

\newcommand{\NAOJ}{\affil{National Astronomical Observatory of Japan, 2-21-1 Osawa, Mitaka, Tokyo, 181-8588, Japan}}

\newcommand{\Nichidai}{\affil{Department of Physics, General Studies, College of Engineering, Nihon University, 1 Nakagawara, Tokusada, Tamuramachi, Koriyama, Fukushima, 963-8642, Japan}}

\newcommand{\NRAO}{\affiliation{National Radio Astronomy Observatory, 520 Edgemont Road, Charlottesville, VA 22903, USA}}

\newcommand{\OAN}{\affiliation{Observatorio Astron\'{o}mico Nacional (IGN), C/Alfonso XII, 3, E-28014 Madrid, Spain}}

\newcommand{\ObsParis}{\affiliation{Sorbonne Universit\'{e}, Observatoire de Paris, Universit\'{e} PSL, CNRS, LERMA, F-75014, Paris, France}}

\newcommand{\Oxford}{\affiliation{Sub-department of Astrophysics, Department of Physics, University of Oxford, Keble Road, Oxford OX1 3RH, UK}}

\newcommand{\Rutgers}{\affiliation{Department of Physics and Astronomy, Rutgers, the State University of New Jersey, 136 Frelinghuysen Road, Piscataway, NJ 08854, USA}}

\newcommand{\STScI}{\affiliation{Space Telescope Science Institute, 3700 San Martin Drive, Baltimore, MD 21218, USA}}

\newcommand{\STScIESA}{\affiliation{AURA for the European Space Agency (ESA), Space Telescope Science Institute, 3700 San Martin Drive, Baltimore, MD 21218, USA}}

\newcommand{\Surrey}{\affiliation{Department of Physics, University of Surrey, Guildford GU2 7XH, UK}}

\newcommand{\Sydney}{\affiliation{Sydney Institute for Astronomy, School of Physics A28, The University of Sydney, NSW 2006, Australia}}

\newcommand{\TAPIR}{\affil{TAPIR, California Institute of Technology, Pasadena, CA 91125, USA}}

\newcommand{\Tamkang}{\affiliation{Department of Physics, Tamkang University, No.151, Yingzhuan Rd., Tamsui Dist., New Taipei City 251301, Taiwan}}

\newcommand{\Toulouse}{\affiliation{Universit\'{e} de Toulouse, UPS-OMP, IRAP, F-31028 Toulouse cedex 4, France}}

\newcommand{\Toledo}{\affiliation{University of Toledo, 2801 W. Bancroft St., Mail Stop 111, Toledo, OH 43606, USA}}

\newcommand{\UChile}{\affiliation{Departamento de Astronom\'{i}a, Universidad de Chile, Camino del Observatorio 1515, Las Condes, Santiago, Chile}}

\newcommand{\UCM}{\affiliation{Departamento de F\'{\i}sica de la Tierra y Astrof\'{\i}sica, Universidad Complutense de Madrid, E-28040, Spain}}

\newcommand{\UCSD}{\affiliation{Center for Astrophysics and Space Sciences, Department of Physics,  University of California, San Diego, 9500 Gilman Drive, La Jolla, CA 92093, USA}}

\newcommand{\ULL}{\affiliation{Departamento de Astrof\'isica, Universidad de La Laguna, Av. del Astrof\'isico Francisco S\'anchez s/n, E-38206, La Laguna, Spain}}

\newcommand{\UMass}{\affiliation{University of Massachusetts—Amherst, 710 North Pleasant Street, Amherst, MA 01003, USA}}

\newcommand{\UVa}{\affiliation{University of Virginia, 530 McCormick Road, Charlottesville, VA 22904, USA}}

\newcommand{\Wyoming}{\affiliation{Department of Physics and Astronomy, University of Wyoming, Laramie, WY 82071, USA}}

\newcommand{\Zurich}{\affiliation{Institute for Computational Science, University of Z\"urich, Winterthurerstrasse 190, 8057 Z\"urich, Switzerland}}

\newcommand{\Lund}{\affiliation{Lund Observatory, Division of Astrophysics, Department of Physics, Lund University, Box 43, SE-221 00 Lund, Sweden}}

\newcommand{\Rad}{\affiliation{Radcliffe Institute for Advanced Studies at Harvard University, 10 Garden Street, Cambridge, MA 02138, U.S.A.}}

\newcommand{\Phillips}
{\affiliation{Phillips Academy, Andover, MA 01810, USA}}

\title{Application of resolved low-$J$ multi-CO line modeling with \texttt{RADEX} to constrain the molecular gas properties in the starburst M82}

\author[0009-0007-2660-7635
]{Valencia Zhang}
\CfA
\Phillips
\author[0000-0002-8760-6157]{Jakob den Brok}
\email{jakob.denbrok@gmail.com}
\CfA
\author[0000-0003-2384-6589]{Qizhou Zhang}
\CfA
\author[0000-0003-4209-1599]{Yu-Hsuan Teng}
%\UCSD
\UMarry
\author[0000-0002-9165-8080]{María~J.~Jiménez-Donaire}
\OAN
\Yebes
\author[0000-0001-9605-780X]{Eric W. Koch}
\CfA
\author[0000-0003-1242-505X]{Antonio Usero}
\OAN 
\author[0000-0003-4793-7880]{Fabian Walter}
\MPIA
\author[0000-0002-3952-8588]{Leindert Boogaard}
\Leiden 
\author[0009-0009-3294-6320]{Craig Yanitski}
\Cologne
\author[0000-0002-1185-2810]{Cosima Eibensteiner}
\altaffiliation{Jansky Fellow of the National Radio Astronomy Observatory}
\NRAO
\author[0000-0003-0583-7363]{Ivana Bešlic}
\ObsParis
\author[0000-0001-8835-218X]{Juan Luis Verbena}
\Cologne

\begin{abstract}
The distribution and physical conditions of molecular gas are closely linked to star formation and the subsequent evolution of galaxies. Emission from carbon monoxide (CO) and its isotopologues traces the bulk of molecular gas and provides constraints on the physical conditions through their line ratios. However, comprehensive understanding on how the particular choice of line modeling approach impacts derived molecular properties remain incomplete. Here, we study the nearby starburst galaxy M82, known for its intense star formation and molecular emission, using the large set of available multi-CO line observations. We present high-resolution (${\sim}85$ pc) emission of seven CO isotopologue lines, including $^{12}$CO, $^{13}$CO, and C$^{18}$O from the $J = 1-0$, $2-1$ and $3-2$ transitions. Using \texttt{RADEX} for radiative transfer modeling, we analyze M82’s molecular properties with (i) a one-zone model and (ii) a variable density model, comparing observed and simulated emissions via a minimum $\chi^2$ analysis. We find that inferred gas conditions—kinetic temperature and density—are consistent across models, with minimal statistical differences. However, due to their low critical densities (${<}10^{4}$ cm$^{-3}$), low-$J$ CO isotopologue lines do not effectively probe higher density gas prevalent in starburst environments like that of M82. Our results further imply that this limitation extends to high-redshift ($z{\gtrapprox}1$) galaxies with similar conditions, where low-$J$ CO lines are inadequate for density constraints. Future studies of extreme star-forming regions like M82 will require higher-$J$ CO lines or alternative molecular tracers with higher critical densities.

\end{abstract}

\keywords{galaxies: ISM -- ISM: molecules -- radio lines: galaxies}

%%%%%%%%%%%%%%%%%%%%%%%%%%%%%%%%%%%%%%%%%%%%%%%%%%%%%%%%%%%%%%%%%%%%%
%
%.     Introduction
%
%%%%%%%%%%%%%%%%%%%%%%%%%%%%%%%%%%%%%%%%%%%%%%%%%%%%%%%%%%%%%%%%%%%%%
\section{Introduction} \label{sec:intro}
Cold, dense clouds of molecular gas within the interstellar medium (ISM) serve as the fundamental reservoirs from which stars form in galaxies. The physical and chemical properties of this gas—such as temperatures, densities, and metallicity—play a crucial role in initiating and regulating the star formation process \citep[see reviews by][]{Omont2007,Kennicutt2012, Schinnerer2024}. These conditions influence the balance between gravitational collapse and thermal support, which determines the efficiency and timescale of star formation.

One common approach to probing the conditions within molecular gas clouds is through observations of rotational transitions of carbon monoxide (CO) and its isotopologues. The assumption of local thermodynamic equilibrium (LTE) is often employed in these studies, as it simplifies the radiative transfer calculations necessary for deriving physical parameters \citep{Wilson2009, Cormier2018,Wange2023}. However, the LTE assumption is frequently inadequate for describing the molecular ISM, where collisional processes and radiative trapping can significantly affect the excitation of molecular lines \citep{Shirley2015}. Consequently, non-LTE modeling approaches have become increasingly important for accurately constraining the conditions of molecular gas. These methods require observations of multiple CO lines to resolve the degeneracies between various physical parameters, such as densities and temperatures \citealt{Finn2021, Teng2022, Roueff2024}).

Utilizing the large set of seven low-$J$ CO isotopologue line observations at the GMC-scale (i.e., ${<}$100\,pc physical resolution), including new SMA data, this study focuses on the bright, nearby starburst galaxy Messier 82 (M82), located at a distance of approximately $D \sim 3.6$ Mpc \citep{Freedman1994}. M82 is one of the closest and most studied starburst galaxies, with an infrared luminosity of approximately $3 \times 10^{10} L_\odot$, making it one of the brightest galaxies observed by the Infrared Astronomical Satellite (IRAS). The central nuclear disk (CND) of M82, spanning about 500\,pc, is a region of intense star formation activity. Compared to the Milky Way’s central molecular zone, M82’s CND is 100 times more luminous \citep{Rieke1980}. Given its extreme star-forming environment, M82 also presents a viable analog for high-redshift galaxies.

Numerous studies have targeted CO and other millimeter (mm) line emissions in M82’s CND to explore the properties of its molecular gas. At a coarser angular resolution ($22''$ or 350\,pc), \citet{Petitpas2000} mapped the 3--2 CO isotopologue lines using the JCMT. Their large-velocity-gradient line modeling suggests the presence of even optically thin $^{12}$CO(1-0) emission in the central region. At a GMC-scale resolution, early work by \cite{Weiss2001} employed high-resolution (1.5 arcsec, 27pc) observations of the 1--0 $^{12}$CO, $^{13}$CO, and C$^{18}$O lines to investigate the galaxy’s star formation-driven dynamics. Their study revealed that the kinetic temperature of the molecular gas is elevated in the central region, where star formation is most active, and decreases towards the outer molecular lobes, where star formation rates are lower. However, with only three CO lines observed, the constraints on physical conditions were limited. Subsequent studies have expanded the dataset of CO line observations in M82, employing both LTE and non-LTE modeling techniques to derive more accurate constraints on the physical conditions of the gas. For example, \cite{Panuzzo2010} utilized CO multi-line observations from Herschel Space Telescope to study the molecular gas for a single beam encompassing the entire central $r{<}1\,\rm kpc$ region of M82. Their work revealed the presence of a very warm molecular gas component at ${\sim}500$~K.

In this study, we aim to further constrain the resolved molecular gas properties in M82 by analyzing high-resolution low-$J$ CO isotopologue observations at 85 pc scale. Specifically, with seven cross-band CO (isotopologue) lines in hand, we set out to benchmark non-LTE modeling of the CO emission with \texttt{RADEX}. We particularly contrast the results derived from assuming a one-zone structure versus a varying density-distribution model to establish how reliable these models are in the context of starburst environments.

This paper is structured as follows: In Section \ref{sec:obs}, we describe the data used in the study. In Section \ref{sec:multiline}, we present non-LTE radiative transfer modeling techniques and the two approaches we employ to describe the emitting gas. In Section \ref{sec:results}, we present the main results, including derived molecular gas properties. We interpret and discuss the results in Section \ref{sec:disc}. Finally, we present the conclusions in Section \ref{sec:conc}.
%%%%%%%%%%%%%%%%%%%%%%%%%%%%%%%%%%%%%%%%%%%%%%%%%%%%%%%%%%%%%%%%%%%%%
%
%.     Data & Observations
%
%%%%%%%%%%%%%%%%%%%%%%%%%%%%%%%%%%%%%%%%%%%%%%%%%%%%%%%%%%%%%%%%%%%%%

\section{Data and Observations} \label{sec:obs}
We obtained observations of seven low $-J$ $^{12}$CO (hereafter CO), $\thirteenco$, and $\coeighteen$ lines with the Submillimeter Array (SMA) and the Northern Extended Millimeter Array (NOEMA). The native angular resolution varies from 2.2$\arcsec$ -- 4.7$\arcsec$, allowing us to resolve the galaxy on a ${\le}85$pc physical scale, which corresponds roughly to the typical GMC scale \citep[see][]{Scoville1987}. We provide an overview of the list of observations as well as their characteristics in Table \ref{table:1}.
% \autoref{table:1}.

\begin{deluxetable*}{ccccccc}
\tablecaption{Summary of CO lines, including telescope, energy transition, emission frequency, and angular and spectral resolution.}

  \label{table:1}
\tablehead{\colhead{Line} & \colhead{Rest Frequency} & \colhead{Angular Scale} & \colhead{Spec. Resolution} & \colhead{$\langle \text{rms} \rangle^a$} & \colhead{Telescope/Reference} & \colhead{Single Dish Available} \\ & [GHz] & [$''$]/[pc] & km\,s$^{-1}$ & [mK]}

\startdata
     C$^{18}$O $J =$ 1--0 & 109.782 & 2.2/40 & 20 & 0.0017 & \multirow{3}{*}{NOEMA} & \textcolor{red}{\ding{56}} \\
      $^{13}$CO $J =$ 1--0 & 110.201 & 2.2/40 & 20 & 0.0022 & &\textcolor{red}{\ding{56}}\\
      CO $J =$ 1--0 & 115.271 & 2.1/38 & 5 & 0.014 & &\textcolor{green}{\ding{52}} \\ \hline
     C$^{18}$O $J =$ 2--1 & 219.560 & 4.7/85 & 5 & 0.0039 & \multirow{4}{*}{SMA} &\textcolor{green}{\ding{52}} \\
      $^{13}$CO $J =$ 2--1 & 220.399 & 4.2/76 &5 & 0.0053&&\textcolor{green}{\ding{52}}\\
      CO $J =$ 2--1 & 230.538 & 3.5/63 & 5 & 0.038&&\textcolor{green}{\ding{52}}\\
     CO $J =$ 3--2 & 345.796 & 0.8/14 & 5 & 0.17& &\textcolor{red}{\ding{56}}\\
%    \bottomrule
%  \end{tabular}
\enddata
\tablecomments{$^{\rm a}$ Channel rms at 5 km\,s$^{-1}$ and working resolution (4.7\arcsec)}
%\end{table*}
\end{deluxetable*}

The NOEMA inteferometric data covers the $J=1{\rightarrow}0$ lines of $^{12}$CO, $^{13}$CO, and C$^{18}$O. The original data covers an area of ${\sim}$25 arcmin\textsuperscript{2}, extending well beyond the CND of M82 out to 7.9 kpc along the major axis and out to 2.9 kpc along the minor axis. The $^{12}$CO(1-0) line has been presented and analyzed in detail in \citet{Krieger2021}. We refer to this publication for details on the data calibration and imaging procedures.

In addition, we present new SMA observations as part of this study, including the $J=2{\rightarrow}1$ lines of  $^{13}$CO, and C$^{18}$O, and both the $J=2{\rightarrow}1$  and $J=3{\rightarrow}2$ for $^{12}$CO. These observations cover only the central CND of M82 and extend to ${\sim}1$\,kpc in radius along the major axis of M82. A detailed description of the SMA data reduction and imaging is provided in J. den Brok et al. (in prep.). We provide a short description of the imaging routine in the Appendix \ref{app:data_cal}.
%%%%%%%%%%%%%%%%%%%%%%%%%%%%%%%%%%%%%%%%%%%%%%%%%%%%%%%%%%%%%%%%%%%%%
%
%.     Results 
%
%%%%%%%%%%%%%%%%%%%%%%%%%%%%%%%%%%%%%%%%%%%%%%%%%%%%%%%%%%%%%%%%%%%%%

\begin{figure*}[]
    \centering
\includegraphics[width=0.9\textwidth]{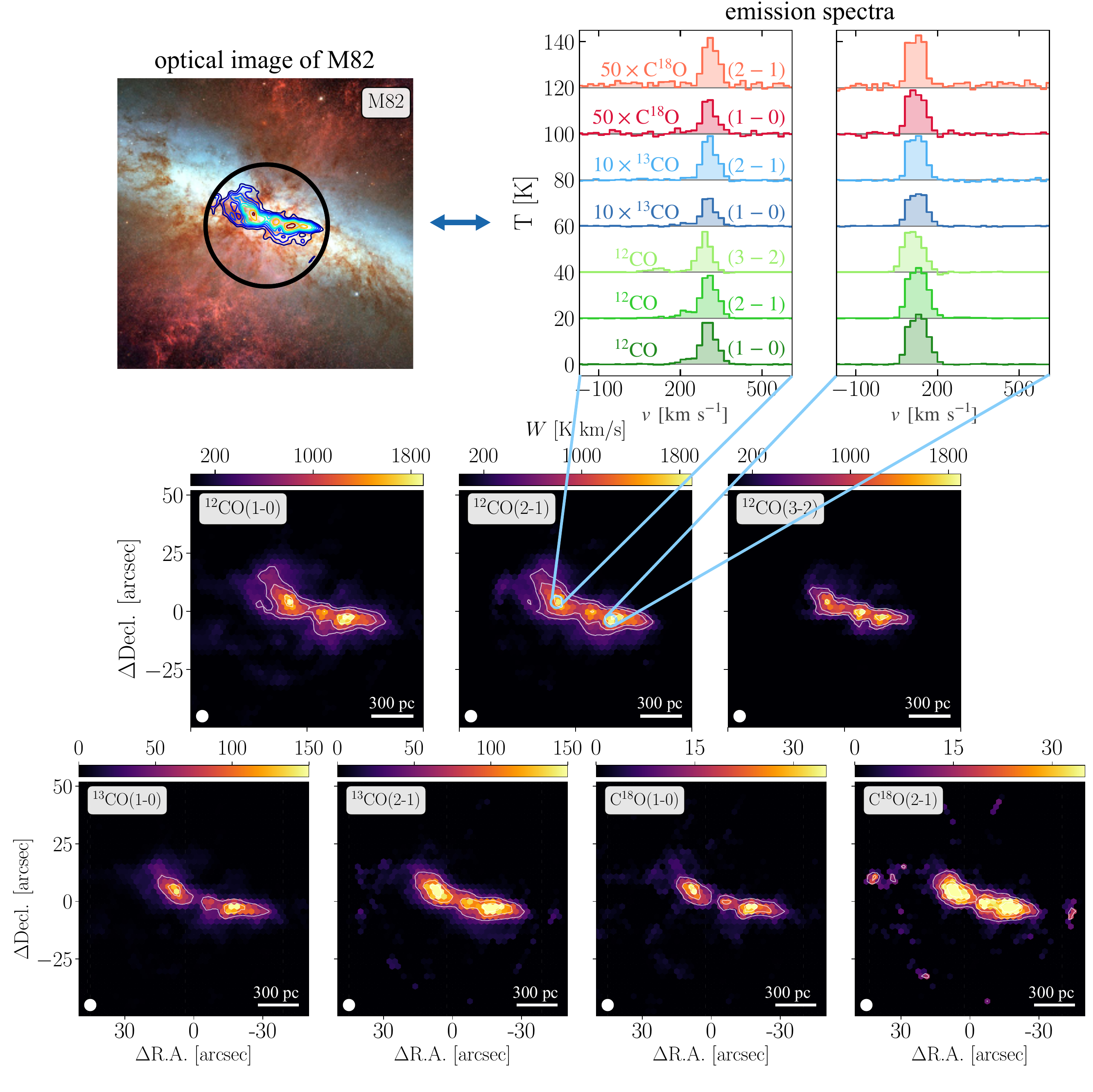}
    \caption{{\bf Optical image, CO Isotopologue Emission and Moment 0 images of M82.} On the top left, we present an optical image of M82. On the top right, we plot emission spectra for two pixels in the center regions of the galaxy. Note that we scale emission from $\thirteenco$ and $\coeighteen$ for visibility purposes. On the lower panel, we present the (1--0) and (2--1) energy transitions from each of the isotopologues \co, \thirteenco, and \coeighteen. \\ Credit for M82 optical Image: NASA, ESA and the Hubble Heritage Team (STScI/AURA). Acknowledgment: J. Gallagher (University of Wisconsin), M. Mountain (STScI) and P. Puxley (NSF).}
    \label{fig:spectramom0}
\end{figure*}
The resulting homogenized data includes the spectral cube and the moment maps derived from it for each line. In Figure \ref{fig:spectramom0}, we present this data. In the top left corner, we present an HST composite optical image of M82. In the top right corner, we plot the spectra of two lines of sight in the central region, with emission from each isotopologue scaled accordingly for visual purposes. Finally, in the lower panels, we plot moment-0 maps from all lines, convolved to 4.7\arcsec and re-sampled on a hexagonal grid. In these maps, bright peaks observed in the CND are clear. These so-called ``knots" of emission are consistent with previous observations (see \cite{Weiss2001}). 

We note that all of the CO isotopologue line data do not include short-spacing corrections (see Table \ref{tab:line_summary}). However, we show in Appendix~\ref{app:data_cal} that in general only  ${<}5\%$ of the flux per line-of-sight is missing, which is well below the assumed flux uncertainty of ${\sim}10\%$. Integrated over the entire field of view, the missing flux for the CO isotopologues still remains at ${\sim}10\%$ for $^{13}$CO and C$^{18}$O (2-1).

\section{Multiline Modeling}\label{sec:multiline}
We use the non-LTE radiative transfer simulation \texttt{RADEX} \citep{vanTak2007} to model observed velocity integrated line intensities (in units K\,km\,s$^{-1}$) from $\co, \thirteenco,$ and $\coeighteen$ under various environmental conditions, which are described by combinations of density of hydrogen ($\hdens$), kinetic temperature ($\tkin$), $^{12}$CO column density per line width ($\coldens/ \Delta v$), abundance ratios ($\xthirteen, \xeighteen$), and -- for the one-zone model -- the beam-filling factor ($\bff$), assumed to be the same for all lines\footnote{We emphasize that the use of a single beam filling factor for the different CO isotopologues is a strong simplification, and can be problematic in the case where selective photodissociation or chemical differentiation leads to a spatial segregation. But stacking results from Galactic clouds \citep{Goldsmith2008,Pety2017} reveal a similar distribution and hence motivate the use of a common beam filling factor.}. Later, we implement a log-normal distribution for $\hdens$ that accounts for a range of densities controlled by distribution width, $\sigma$. We note that incorporating CO isotopologue abundances into \texttt{RADEX} modeling has been shown to improve the recovery of molecular gas physical conditions \citep{Tunnard2016}.

\texttt{RADEX} assumes homogeneous gas and solves the radiative transfer equations to find a converged solution for the level population \citep{vanTak2007}. We use molecular data files from the Leiden Atomic and Molecular Database \citep{schoierRADEX2005}, and we assume the CO isotopologues only collide with $\hydrogen$ because $\hydrogen$ is much more abundant than any other molecule ($[\co / \hydrogen] \sim 10^{-4}$; \citealt{Draine2011}).

We consider two models to describe emitting molecular gas, which are described in more details in the following subsections. First, following the common approach in the literature, we assume that the conditions of emitting gas in each line of sight are uniform (i.e. one-zone model). Next, to better describe the gas, we implement a log-normal density model, where $\hdens$ follows a log-normal distribution. 

For ${\sim}20\%$ of the sightlines, we notice multiple gas components in the emission spectra. This is expected, as M82 is nearly edge-on. Because the multi-component behavior is not dominant in the central region, we assume a singular component for all sightlines. We note that the solution will be biased towards the brighter component in the case of multiple components, as shown in \citet{Teng2023}. 

\begin{table*}
\caption{Summary of Gas Phase Initial Conditions for Both Models}
\label{table:radex}
\centering
\setlength{\extrarowheight}{3pt}
\begin{tabular*}{\textwidth}{@{\extracolsep{\fill}}cccc}
\toprule
 Model Prescription & Parameter & Range & Stepsize \\
\midrule
\multirow{5}{*}{Both Models} 
& Kinetic temperature, $\log_{10} (\mathrm{T}_{\mathrm{kin}}\, [\mathrm{K}])$ & 0.6 -- 2.2 & 0.1 dex \\
& CO column density, $\log_{10}(N_{\mathrm{CO}} \, [\mathrm{cm}^{-2}])$ & 16.0 -- 19.0 & 0.2 dex \\
& Abundance ratio, $\mathrm{X}_{12/13}$ & 10 -- 200 & 10 \\
& Abundance ratio, $\mathrm{X}_{13/18}$ & 2 -- 20 & 1.5 \\
& Line width$^{a}$, $\Delta v \, [\mathrm{km} \mathrm{s}^{-1}]$ & 10 & ... \\
\midrule
\multirow{2}{*}{One-zone}& Hydrogen volume density, $\log_{10} (n_{\mathrm{H}_{2}}\, [\mathrm{cm}^{-3}])$ & 2.0 -- 5.0 & 0.2 dex \\
& Beam-filling factor, $\log_{10} (\bff)$ & ${-}2$ -- 0 & 0.2 dex\\
\midrule
\multirow{2}{*}{Log-normal}& Mean hydrogen volume density, $\log_{10} (n_{0,\mathrm{H}_{2}}\, [\mathrm{cm}^{-3}])$ & 2.0 -- 5.0 & 0.2 dex \\
 & Density distribution width, $\width$ & 0.2 -- 1 & 0.1 \\
\bottomrule
\end{tabular*}

{\vspace{-5mm} \begin{flushleft} NOTE -- {($a$) In essence, \texttt{RADEX} models the line intensities based on the column density per linewidth ($N_{\rm CO}/\Delta v$). We implement this by allowing a range of column densities but fixing the line width. The derived intensities are consistent upon normalization by the line width.} \end{flushleft}}

\end{table*}

\subsection{One-zone Model}
Assuming that the gas in each sightline is uniform, the emission can be described by single values of $\hdens, \tkin, \coldens, \xthirteen, \xeighteen$ with the assumption of an identical $\bff$ for all lines. We run \texttt{RADEX} by iterating through the 6 parameters {following} the ranges and stepsizes outlined in Table \ref{table:radex}. {The radiative transfer calculations in \texttt{RADEX} depend solely on the $\coldens/\Delta v$ ratio, resulting in a degeneracy between $\coldens$ and $\Delta v$ \citep{vanTak2007}. The \texttt{RADEX} calculations allow us to estimate $\coldens/\Delta v$, which we then multiply by the observed CO(1–0) line width for each pixel to derive the true $\coldens$ and thereby avoid this degeneracy. For the model grid,} we assume a fixed $\Delta v = 10$ km\,s$^{-1}$ for both models.\footnote{{We note that this choice is arbitrary, since the results are identical upon normalization by the line width}. } These parameters generate a six-dimensional (6D) grid, amounting to $\sim$12 million points. Our modeling approach and parameter set-up follow those in \cite{Teng2022}. 

\subsection{Log-normal Density Model}
Theoretical and observational studies have shown that molecular clouds have a range of $\hydrogen$ densities from diffuse to dense ones, and cannot necessarily be described by a single density as we assumed above \citep{Nishimura2019}. In effort to describe the molecular gas more physically and explore the effects of a more sophisticated model, we implement a log-normal distribution for the density of $\hydrogen$. The log-normal distribution assumption follows from the literature, which analyzed high resolution observations of molecular cloud in the CO emission \citep{Hughes2013}. Other literature also employed simulations and found a similar log-normal result (see \cite{MacLow2004, Elmegreen2002, Vazquez1994}). Following formulations modeling the density of a cloud, we adopt the probability density function (PDF) method used in \cite{Leroy2017}. See more details in Appendix \ref{app: lognormal}. 

In short, the final intensities are the weighted sum over \texttt{RADEX} one-zone modeled intensities for different volume densities, but for a fixed temperature, abundance, and column density per unit line width. For this calculation, we sum over H$_2$ volume densities ranging from 10$^{-1}$ to $10^7$ cm$^{-3}$ in steps of 0.2 dex. Under this model, the final set of parameters governing each line of sight therefore are $\hdensLog, \tkin, \coldens, \xthirteen, \xeighteen$ and $\sigma$, the width of the log-normal density distribution. We note, that strictly speaking, $n_{0,\rm H_2}$ does not correspond to the average volume density introduced for the one-zone model ($n_{\rm H_2}$). For the log-normal model, $n_{0,\rm H_2}$ more precisely corresponds to the mean volume-weighted density. For modeling purposes, we assume a beam filling factor $\bff$ of 1 for all density layers to limit the degrees of freedom.
We emphasize that this is a simplification, and in certain extragalactic environments, there might be indeed regions within a $4''$ beam that do not exhibit the presence of any (CO-rich) molecular gas, hence implying the need for a filling factor. However, in the case of the central region of M82, we find a beam filling factor from the one-zone model that is quite large ($\sim$0.5). 
The exact value will be part of a future study where we include more mm molecular line tracers.\footnote{A key simplification in our log-normal density model prescription lies in modeling the beam-averaged intensity as a density-weighted sum. In principle, each density layer could have a distinct beam filling factor, which likely depends on the local volume density. A more detailed modeling approach could introduce a \emph{global} beam filling factor to scale the modeled intensities in equation \ref{eq:chi2}. However, in this study, we refrain from adding this additional free parameter due to the limited number of available lines. While this simplification might introduce a bias, it does not impact the main conclusion: even with a global beam filling factor, the density cannot be reliably constrained above a volume density of $10^4\,$cm$^{-3}$. } The ranges through which we iterate can be found in Table \ref{table:radex}. {More details about the implementation of this density distribution can be found in Appendix~\ref{app: lognormal}.}

\subsection{Fitting emission models to observations}\label{subsec:fitting}
We evaluate the goodness of fit of each parameter set $\theta = (\hdens, \tkin, \coldens, \xthirteen, \xeighteen, \bff)$ or $\sigma$ (depending on model) by calculating $\chi ^2$. To remove nonphysical parameter sets from consideration, we set a prior on the line-of-sight path length ($\ell_{\mathrm{los}}$), following \citet{Teng2022}, with (where the beam filling factor term is only considered for the one-zone model)
\begin{equation}
\ell_{\text{los}} = N_{\text{CO}} \left( \sqrt{\Phi_{\text{bf}}} \, n_{\text{H}_2} \, x_{\text{CO}} \right)^{-1} < 500 \, \text{pc},
\end{equation}
where 500pc is based on the width of M82's central region.

Assuming a multivariate Gaussian probability distribution, we convert the $\chi ^2$ for each grid point to a probability. Finally, we generate marginalized 1D and 2D likelihood distributions (called PDFs for 1Ds) for every parameter and pair of parameters. These are generated by summing the probability distribution over the full range of parameter(s) {except} for those of interest. The resulting ``1DMax solution" parameter is then selected by determining the highest 1D likelihood for each parameter. More details can be found in Appendix \ref{app: likelihood}.

\section{Results} \label{sec:results}
In combination, our homogenized dataset consists of 95 sightlines for which all six CO isotopologues lines are detected significantly at S/N${>}$5. {We note that in this analysis, we exclude the CO (3-2) line due to insignificant emission. More discussions on the inclusion of the $J = 3-2$ line can be found in Subsection \ref{discussion: lognormal}.} In the following analysis, we obtain constraints for these 95 sightlines based on our \texttt{RADEX} line modeling. At half beam sampling, given an oversampling factor of ${\sim}4$, this means we have ${\sim}25$ independent measurements.

\begin{figure*}[]
    \centering
\includegraphics[width=1\textwidth]{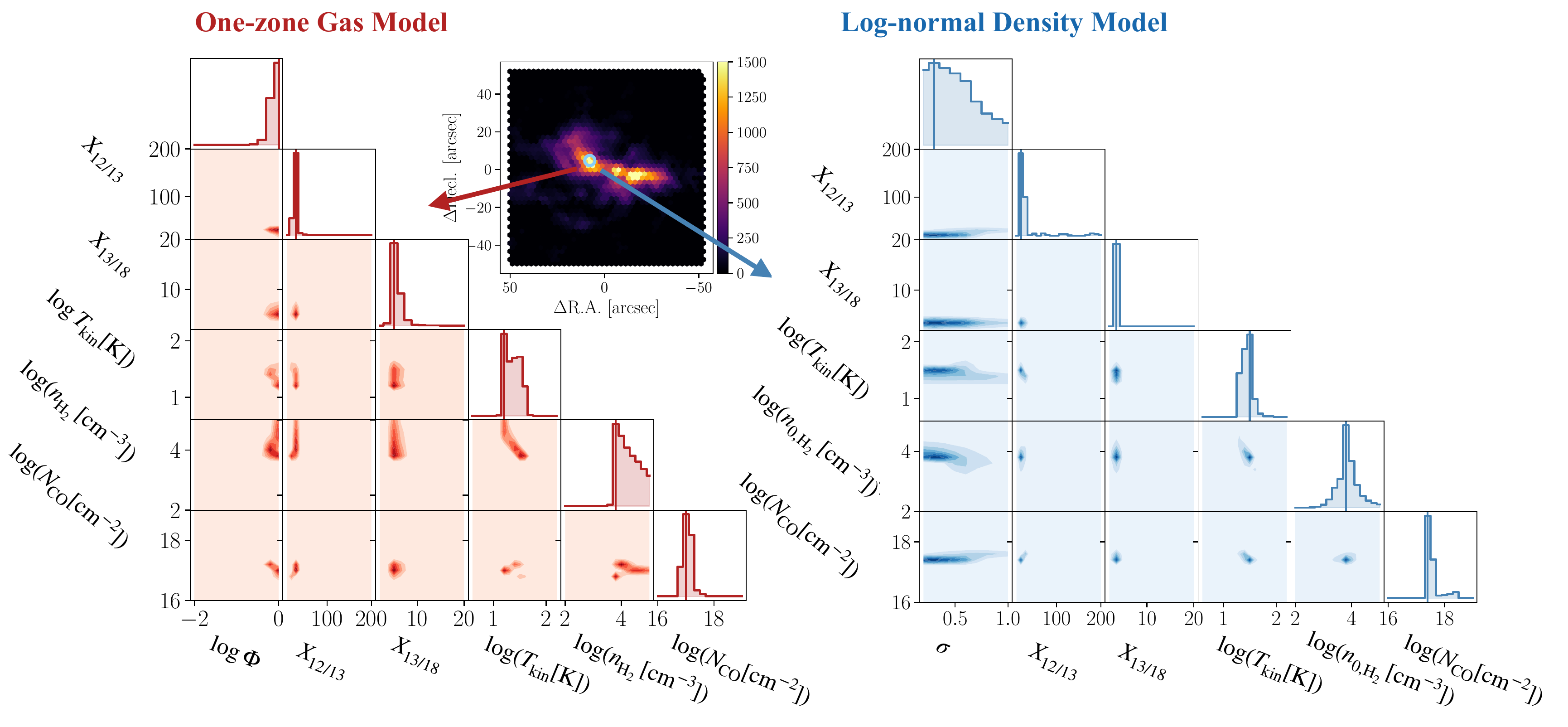}
    \caption{{\bf Marginalized 1D and 2D likelihood distributions from a bright central pixel.} The red (left) corner plot is derived from our one-zone model, and the blue (right) corner plot is derived from our log-normal model. The map in the middle depicts the integrated intensity with one sightline circled from which the two corner plots are generated.}
    \label{fig:bothcorner}
\end{figure*}

We present the results from our multiline modeling in Figure \ref{fig:bothcorner}, where we use corner plots to illustrate the PDFs spanning the parameter space. We present corner plots for a single, representative, bright pixel (note that we obtain such corner plots for each individual sightline). The corner plots depict both the 1D and 2D PDFs. The vertical line on the PDFs along the 1D distributions is the 1DMax solution. The statistics of the 1DMax solutions are listed in Table \ref{tab:line_summary}. We see the 1D likelihoods of $\bff, \xthirteen, \xeighteen$ and $\coldens$ for the one-zone model are well constrained. For the log-normal model, $\xthirteen, \xeighteen$, and $\coldens$ are very tightly constrained as demonstrated by very narrow PDFs. In both models, %the $\tkin$ is constrained moderately well, with both PDFs indicating some form of "pile-up" in the probability. T
the PDF is narrow, indicating that $\tkin$ is reasonably constrained. In contrast, the $\hdens$ parameter is more loosely constrained. For most corner plots, the 1D PDF of $\sigma$ is wide, demonstrating an inability to tightly constrain the parameter. This is discussed further in Section \ref{discussion: lognormal}.

\subsection{Comparison between one-zone and log-normal derived properties}\label{subsec: compareresults}
\begin{figure*}[]
    \centering
\includegraphics[width=0.9\textwidth]{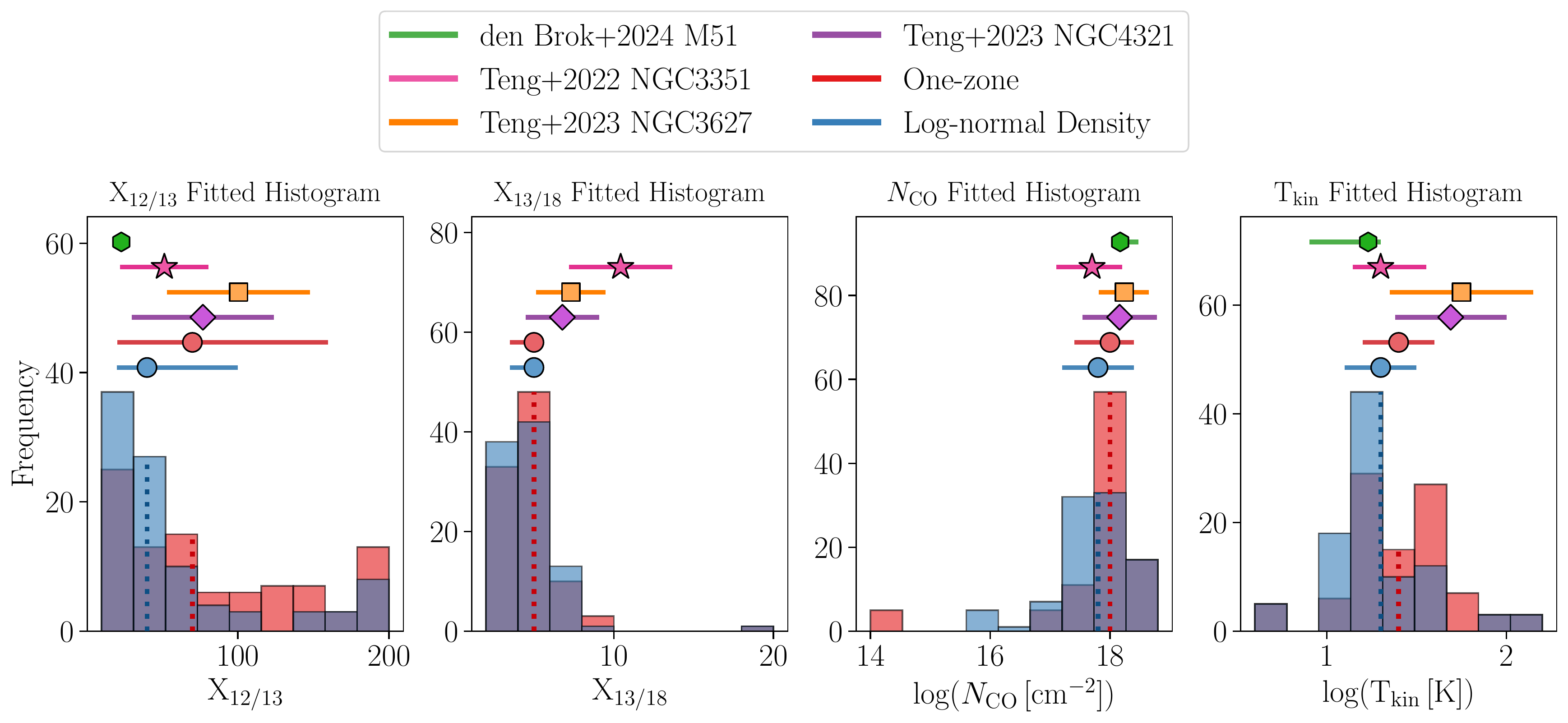}
    \caption{{\bf Histogram of 1DMax fitted parameters for $\xthirteen$, $\xeighteen$, $\coldens$, and $\tkin$ from both models.} We plot histograms of 1DMax fitted parameters, with the one-zone model (red) and the log-normal density model  (blue). The vertical dotted line is the median of the 1DMax values. In the horizontal markers above, we plot the weighted 16$^{\rm th}$-to-84$^{\rm th}$ percentile range from our results, from results found in \cite{Teng2023} for nearby barred galaxies, and from \citet{denBrok2024_M51} for M51.}
    \label{fig:overlayhist}
\end{figure*}

Our two approaches to modeling emitting gas both yield 1DMax solutions for all 6 parameters over the center of the galaxy. In Figure \ref{fig:overlayhist}, we plot the binned fits of parameters shared between the models, which include $\xthirteen, \xeighteen, \coldens$, and $\tkin$. We bin the solutions of the 95 sightlines for which we have significant CO isotopologue emission (excluding $\co$ (3-2)). We see similar behavior when including the $\co$ 3-2 line, but then being limited to only 34 pixels. We make two notes: first, because we consider the effect of $\bff$ in the one-zone model, the 1DMax $\coldens$ value constrained is the sub-beam value, i.e the $\coldens$ in the smaller beam described by $\bff$. In Figure \ref{fig:overlayhist}, we convert from the sub-beam value to the full-beam value by multiplying by $\bff$. Second, the $\hdens$ values from the two models are not directly comparable because for the log-normal model, $\hdens$ is the average density rather than the singular density. It is still instructive, however, to consider their differences. 

Figure \ref{fig:overlayhist} includes the distribution of the 1DMax solutions. The dotted vertical lines are the medians of the 1DMax solutions over all pixels, and the horizontal markers at the top of the plots indicate the weighted 16$^{\rm th}$-to-84$^{\rm th}$ percentile range. Above the fits, we show values found in the centers of NGC3627, NGC4321, and NGC3351 which are nearby barred galaxies \citep{Teng2022, Teng2023}, and the M51 galaxy, which is a nearby grand design spiral galaxy \citep{denBrok2024_M51}. The $\xeighteen$ parameter was not constrained for in \citet{denBrok2024_M51} and thus is left empty in the figure. We see that the two models generally agree for the medians of the $\xeighteen$, $\coldens$, and $\tkin$ parameters.  We see that the median of the one-zone $\xthirteen$ values is higher -- but within the margin of the scatter -- than that of the log-normal model. This likely reflects the fact that in the case of the one-zone model, sub-beam density variations and the resulting CO isotopologue emission are compensated by higher CO isotopologue abundance. 
In comparison to values found for other galaxies, we see generally more extreme conditions in the barred galaxies and less extreme conditions in M51. 

We performed the Kolmogorov-Smirnov (KS) test to quantitatively evaluate the similarity between the results. The p-values from the KS test are listed in Table \ref{tab:line_summary}. For 3 of the 5 parameters shared by the models, the p-value was greater than 0.05, indicating that the difference between the results is not statistically significant. For the two parameters where the p-value was found to be $\ll 0.05$, namely $\tkin$ and $\coldens$, the medians were verified to be within 0.1 dex. 

\subsection{Optical Depth Calculation}
\label{sec:optdepth}
For each sightline, we derive the line center optical depth of the 1-0 transition of $\co, \thirteenco,$ and $\coeighteen$ as follows: for each pixel, we marginalize all modeled optical depths over the whole grid to obtain a 1DMax estimate. We note that in the subsequent section, we work with the optical depths derived from the one-zone model. A similar calculation for the log-normal density model is more ambiguous, since each layer will have a different opacity. 

\begin{deluxetable*}{llccccccc}
    \tablecaption{Statistics for 1DMax solutions across all pixels. We list the median, mean, and standard deviation of the derived 1DMax parameters. For the log-normal model, the volume density refers to the mean volume density, $n_{0,\rm H_2}$.  \label{tab:line_summary}}

    \tablehead{\colhead{Model}&&\colhead{$\log_{10}\left(n_{\rm H_2}\right)$}&\colhead{$\log_{10}\left(T_{\rm kin}\right)$}&\colhead{$\log_{10}\left(N_{\rm CO}\right)$} & \colhead{$X_{12/13}$}&\colhead{$X_{13/18}$}&\colhead{$\Phi$}&\colhead{$\sigma$}\\  & & $\left[\rm cm^{-3}\right]$& $\left[\rm K\right]$ & $\left[\rm cm^{-2}\right]$& &&&$[\rm dex]$ }
\startdata 
         \multirow{3}{*}{One-zone}& Median & 3.71 & 1.4 & 18.2 & 60.0 & 5.0 & 0.4 & --\\
         & Mean & 3.4 & 1.65 & 18.46 & 74.84 & 5.3 & 0.59 & --\\
         & Std. Dev. & 0.42 & 0.32 & 0.43 & 42.51 & 2.71 & 0.27 & --\\ \hline
         \multirow{3}{*}{Log-normal}& Median & 4.2 & 1.3 & 17.80 & 60.0 & 3.5 & -- & 0.4\\
         & Mean & 4.3 & 1.33 & 17.93 & 70.21 & 3.89 & -- & 0.50\\
         & Std. Dev. & 0.56 & 0.19 & 0.44 & 46.07 & 2.63 & --&0.35\\ \hline
         Comparison$^{\rm a}$ & p-value & 0.095 & ${\ll}0.05$& ${\ll}0.05$& 0.099& 0.99 & -- & --
\enddata
\tablecomments{(a) We compare the two distributions using the KS statistic. 
}
\end{deluxetable*}

\begin{figure*}
    \centering
\includegraphics[width=0.8\textwidth]{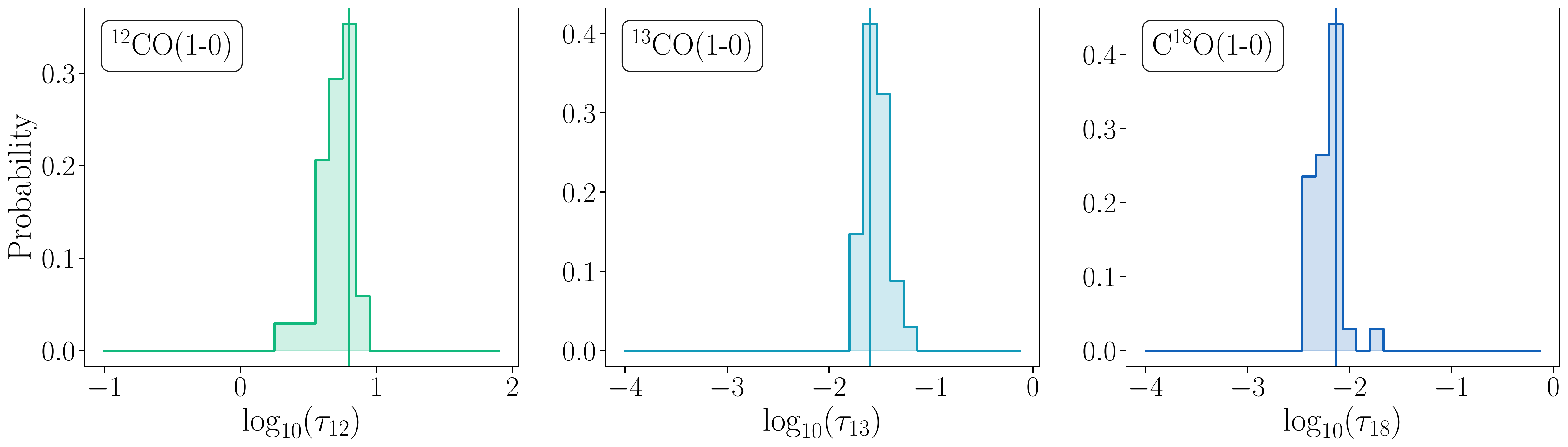}
    \caption{{\bf Computed optical depths for the 1-0 transition for $\co, \thirteenco$, and $\coeighteen$.} We bin the optical depths computed from the 1DMax physical conditions for the one-zone model. As expected, we find optically thick $^{12} \co$ and optically thin $\thirteenco$ and $\coeighteen$.}
    \label{fig:opticaldepth}
\end{figure*}

In Figure \ref{fig:opticaldepth}, we bin the 1DMax solutions from each isotopologue. We refer to relative frequency as probability. We find (in the leftmost panel) that the $\co$ emission is optically thick, as $\tau_{12} > 1$ for all sightlines. We note that the derived values also align with those found for the centers of other nearby galaxies, including NGC 3351, 3627, and 4321, using a similar modeling approach \citep{Teng2022,Teng2023}. In contrast, we note that \citet{Petitpas2000} report lower optical depths (around 0.3--3 for $^{12}$CO(1-0)). However, these values correspond to measurements at 22\arcsec and also depend on the range of assumptions of kinetic temperatures and the CO isotopologue abundances. We note that their optical depths are also above unity if we select their model results which used values for kinetic temperatures (30\,K) and $^{13}$CO/$^{12}$CO abundance ratio (70) that are similar to what we report.

We find that the $\thirteenco$ and $\coeighteen$ 1-0 emission is exclusively optically thin with $\langle\log_{10}\tau_{13}\rangle= -1.5 \pm 0.1$ and $\langle\log_{10}\tau_{18}\rangle=-2.2\pm 0.1$. This aligns well with our expectations of optically thin $\thirteenco$ and $\coeighteen$ emission based on the measured (2-1)/(1-0) excitation ratios. The mean integrated intensity ratio over the 95 sightlines for $^{13}$CO(2-1)/(1-0) is 1.62 and for C$^{18}$O(2-1)/(1-0) is 2.03. Such high ratios are expected in case both the 1-0 and 2-1 transitions remain optically thin and therefore are consistent with our results from the CO line modeling.

%%%%%%%%%%%%%%%%%%%%%%%%%%%%%%%%%%%%%%%%%%%%%%%%%%%%%%%%%%%%%%%%%%%%%
%
%.     Discussion 
%
%%%%%%%%%%%%%%%%%%%%%%%%%%%%%%%%%%%%%%%%%%%%%%%%%%%%%%%%%%%%%%%%%%%%%
\section{Discussion} \label{sec:disc}
\subsection{Limitations of CO isotopologue model applications}\label{discussion: lognormal}

As discussed in Section \ref{subsec: compareresults}, we find that the constraints between the one-zone and log-normal model results do not significantly differ for the volume density, $\hdens$, of H$_2$ and the CO isotopologue abundances.

In principle, we expect the log-normal distribution to more closely trace the true physical conditions of the molecular gas since it adopts a range of densities rather than a single value, as in the one-zone model. Inspecting the individual corner plots, however, we see that there is a sharp cut off  for the $\hdens$ PDF roughly near $10^4$ cm$^{-3}$, with the values above being weakly or not constrained. Similarly, for the log-normal distribution, while the volume density seems to be constrained around $10^4$ cm$^{-3}$, we only get a loose constraint on a wide $\sigma$ PDFs. This implies that low $-J$ CO lines are unable to effectively trace higher gas densities that are most likely present in starburst galaxies. This is consistent with expectations based on the critical density and optical depth effects \citep{Shirley2015}. Indeed, the CO isotopologues used in the study have critical densities on the order of $10 ^3 - 10^4$\,cm$^{-3}$, which decrease further by roughly another order of magnitude in the optically thick gas. 

Finally, we also note that the two model results show similar distributions of derived molecular gas properties (see \autoref{fig:overlayhist}), despite the simplified assumptions employed for both prescriptions. This suggests that the major limitation is inherent to the general low-$J$ CO line ratio modeling approach and only to a lesser degree to simplified assumptions, such as a uniform gas layer (for the one-zone model), or no global beam filling factor for the log-normal density model.

There are several indications that support the presence of high density ($n_{\rm H_2}{>}10^4$\,cm$^{-3}$) and warmer ($T_{\rm kin}{>}50\,$K) molecular gas in the central region of M82:(i) the molecular gas in starbursts is observed to show densities that are much higher than those found in typical disk galaxies  \citep{Jackson1995,Iono2007}. (ii) observations also suggest that the molecular gas in starbursts is warmer \citep{Bradford2003,Rangwala2011}. Such high temperatures and densities are consistent with the high (2-1)/(1-0) line ratios observed for the CO isotopologues in M82 (see Section \ref{sec:optdepth}).

Studies using different molecular gas tracers than the low-$J$ CO isotopologues indeed report higher density and temperature gas in the central region of M82. For instance, \citet{Panuzzo2010} rely on higher-$J$ CO lines observed with \textit{Herschel}, including the CO(13-12) and report temperatures in the range of 400--800\,K, an order of magnitude higher than the average temperatures we find using the CO line modeling approach. Using the formaldehyde lines as a thermometer, \citet{Muehle2007} also report temperatures of around 200\,K for the central region. Previous studies also found densities above 10$^4$\,cm$^{-3}$. For instance, using higher critical density tracers, including HCN, CS, and HCO$^{+}$, \citet{Naylor2010} report average molecular gas densities not in excess of 10$^{4.4}$\,cm$^{-3}$.

Therefore, our results suggest that even with sophisticated modeling approaches incorporating numerous low-$J$ CO isotopologue lines, it remains crucial to include higher excitation energy lines (e.g., higher-$J$ CO lines beyond the 2–1 transition) and high critical density transitions (e.g., multiple HCN transitions) to achieve robust constraints in extreme environments, such as the central region of M82. 
{Indeed, previous Galactic and extragalactic studies focusing on higher-$J$ lines or lines of higher critical density do recover higher volume densities. For instance, studies using \textit{Herschel} SPIRE/FTS and PACS spectra covering higher-$J$ lines of up to $J_{\rm up}{=}30$, do recover higher temperatures (${>}100\,$K) and volume densities (${\ge}10^{5}$\,cm$^{-3}$) in local active galaxies \citep{PereiraSantaella2013, PereiraSantaella2014, Mashian2015} and nearby starbursts \citep[NGC 6240 and Arp 193;][]{Papadopoulos2014}. Similarly, studies including dense gas tracers, such as the rotational HCN and HCO$^+$ lines, also recover and constrain the volume densities and temperatures of a hot (${>}100\,$K) and dense ($>10^{5}$\,cm$^{-3}$) molecular gas component in extreme conditions found in ULIRGs \citep[e.g.,][]{Mashian2015,Imanishi2023_Ulirg,Imanishi2023} or other local starbursts \citep{Krips2008,Greve2009}. These prior findings reinforce our conclusion that relying on low-$J$ CO isotopologue transitions alone can lead to an incomplete characterization of molecular gas conditions in extreme environments, such as the central region of M82, and }underscore the importance of expanding the range of molecular tracers to capture the varied physical conditions in starbursts.

The intense star formation and high molecular gas densities in M82 serve as a compelling nearby analog to the molecular gas conditions in high-redshift galaxies \citep[$z\gtrapprox1$; e.g.,][]{Weiss2005, Casey2014}. Our findings underscore the need for caution when interpreting low-$J$ CO lines in studies of high-$z$ systems. Analyses relying solely on low-$J$ transitions risk mis-characterizing the density and excitation properties of molecular gas in these environments. To improve the accuracy and diagnostic power of molecular line studies, future work should prioritize the inclusion of multi-line, multi-tracer modeling and higher-density tracers, particularly in starburst galaxies, both locally and at high redshift.

%%%%%%%%%%%%%%%%%%%%%%%%%%%%%%%%%%%%%%%%%%%%%%%%%%%%%%%%%%%%%%%%%%%%%
%
%.     Conclusion
%
%%%%%%%%%%%%%%%%%%%%%%%%%%%%%%%%%%%%%%%%%%%%%%%%%%%%%%%%%%%%%%%%%%%%%
\section{Conclusion} \label{sec:conc}
We present analysis of high resolution ($85$\,pc physical scale) observations of CO isotopologues from seven different low-$J$ transitions across the center of M82. We model emitting molecular gas under two assumptions: one-zone and log-normal density distributions. We use these models together with a non-LTE radiative transfer simulation from \texttt{RADEX} and likelihood analysis to constrain the physical conditions of the molecular gas. We obtain solutions for 95 sightlines within the central ($r{<}1\,$kpc) region of M82. Our main findings are as follows:
\begin{enumerate}
    \item The derived parameters from the two models, particularly the H$_2$ volume density and relative CO isotopologue abundances, are not statistically significantly different. Moreover, models do not tightly constrain the $\sigma$ parameter, which describes the width of the log-normal density distribution.
    \item The derived optical depths of the 1-0 transition of $^{12}$CO are optically thick and 
    are optically thin for $^{13}$CO and C$^{18}$O for all sightlines. In particular, the optically thin emission for the CO isotopologue lines is expected as their (2-1)/(1-0) line ratios are above unity. 
    \item While higher-$J$ lines are critical to constrain the conditions in particular environments of high densities and temperatures, we find that the CO 3-2 line alone does significantly improve the estimate of molecular gas conditions for most sightlines. Overall, we find that the effectiveness of low-$J$ CO isotopologue line modeling might be limited in the case of extreme environments found in starbursts, where gas densities exceed the critical density significantly and the temperatures exceed the excitation temperatures of the lines.
\end{enumerate}
For future work,  higher$-J$ CO lines as well as emission from other molecules with higher critical densities, such as HCN and HCO$^+$ are needed for robust non-LTE analyses of molecular gas.

\section*{acknowledgments}
{We thank the anonymous referee for going carefully through the paper  and appreciate their helpful and constructive comments that improved the clarity of this paper.} We use data from the Submillimeter Array which  is a joint project between the Smithsonian Astrophysical Observatory and the Academia Sinica Institute of Astronomy and Astrophysics and is funded by the Smithsonian Institution and the Academia Sinica. {We recognize that Maunakea is a culturally important site for the indigenous Hawaiian people; we are privileged to study the cosmos from its summit.} This work is also based on observations carried out  under project IDs
w18by and 107-19 with the IRAM NOEMA Interferometer and 30m telescope. IRAM is supported by INSU/CNRS (France), MPG (Germany) and IGN (Spain). JdB and EWK acknowledge support from the Smithsonian Institution as Submillimeter Array (SMA) Fellows. MJJD and AU acknowledge support from the Spanish grant PID2022-138560NB-I00, funded by MCIN/AEI/10.13039/501100011033/FEDER, EU.
%\end{acknowledgments}

\vspace{5mm}
\facilities{Submillimeter Array (\textit{SMA}), Institut de radioastronomie millimétrique (\textit{IRAM}) 30m and Northern Extended Millimeter Array (\textit{NOEMA}) telescopes}

\software{astropy \citep{2013A&A...558A..33A,2018AJ....156..123A},  
numpy \citep{harris2020array}, CASA \citep{CASA}, pyuvdata \citep{Hazelton2017}
          }

\appendix

\section{Interferometric SMA data calibration and imaging}
\label{app:data_cal}

The SMA observations of the 2-1 transition of $^{12}$CO, $^{13}$CO, and C$^{18}$O were carried out as part of the 2022 SMA Interferometry School (program ID: 2021B-S059; PIs: J. den Brok, C. Eibensteiner, I. Beslic) and a dedicated wide-band SMA survey covering the 200--300 GHz frequency range (program IDs: 2022B-S034, 2023B-S033; PI: J. den Brok). The data calibration and imaging procedures for these observations will be detailed in J. den Brok et al. (in prep.). For completeness, we provide a brief summary of the data reduction and imaging steps here.

\subsection{SMA Observations and Data Calibration}

The data presented in this project are a combination of observations from three separate tracks carried out on 2022 January 5, 2023 January 24, and 2023 January 25. The 2022 observations, which were part of the SMA Interferometry School program, were obtained in the compact configuration, while the 2023 observations were conducted with the SMA in the subcompact configuration. All tracks were observed under favorable weather conditions, with a nightly average $\langle\rm pwv\rangle{\lesssim}4$,mm.

Throughout the observing sessions, we used a set of flux, gain, and bandpass calibrators, selected based on their availability. For all tracks, Uranus was used as the flux calibrator. The gain calibrators, 0958+655 and 0841+708, were observed, as both were sufficiently bright, with a flux density of ${>}$0.5 Jy, allowing for efficient calibration with scans of $<$ 2 min each. For bandpass calibration, either 0319+415 or 3c84 was selected.

The calibration of the SMA data itself was carried out using the Common Astronomy Software Applications (CASA; \citealt{CASA}) package. For this, the raw SMA data were converted to CASA-readable measurement set (MS) files using the pyuvdata \citep{Hazelton2017} utilities. We then employed the SMA data calibration scripts\footnote{\url{https://github.com/Smithsonian/sma-data-reduction}} to perform the bandpass, gain, and flux calibration.

\subsection{SMA Imaging using the PHANGS Imaging Pipeline}

Using the calibrated CASA-readable measurement sets, we use the \textit{PHANGS-ALMA imaging pipeline} \citep{Leroy2021}\footnote{\url{https://github.com/akleroy/phangs_imaging_scripts/}} to perform the line imaging. In order for the PHANGS pipeline to work for SMA data, we have to mimic the ALMA format by adding scan intents to the measurement set table. The final imaging is performed using the \texttt{tclean} task from casa. The pipeline performs the continuum subtraction and cleaning automatically. We refer to \citet{Leroy2021} for a detailed description of the functionality and individual steps of the pipeline.
We note that we do not reweight the visibilities as the SMA's weights are well-defined based on system temperature measurements made throughout the observations.

\subsection{Assessing the degree of missing flux due to missing short-spacing correction}

\begin{figure*}
    \centering
\includegraphics[width=0.7\textwidth]{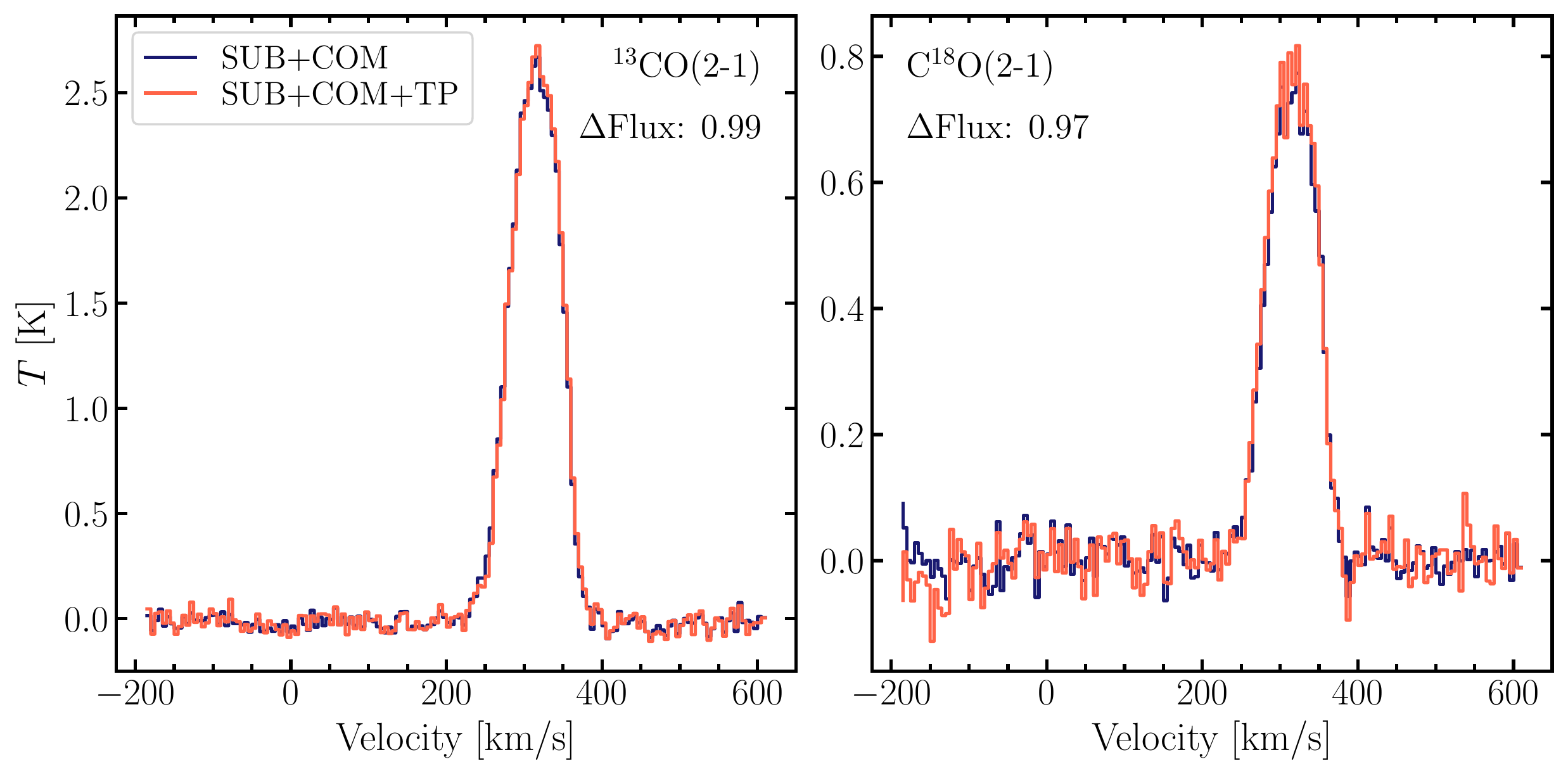}
    \caption{{\bf Comparison of interferometric and short-spacing corrected data along a line of sight in M82.} We illustrate the missing flux for an arbitrary sightline in the center of M82, resolved at 85\,pc  for $^{13}$CO(2-1) on the left and C$^{18}$O(2-1) on the right. Blue shows the interferometric data only and red shows the short-spacing corrected data using DDT IRAM 30m observations. We note that the difference in flux, $\Delta$Flux, quotient of interferometric only over short-spacing corrected data over the spectrum, is indicated in each panel and is less than 3\% of the total integrated intensity.  }
    \label{fig:missing_flux}
\end{figure*}

For the NOEMA 3mm $^{13}$CO and C$^{18}$O observations, we lack short-spacing corrected observations. Therefore, for consistency, we also work with the 1.3mm observations before being short-spacing corrected. Using DDT 30m observations (E20-02; PIs: J. den Brok, C. Eibensteiner, I. Beslic), we can assess to what degree the SMA data are affected by the lack of the shortest baselines. In \autoref{fig:missing_flux}, we provide a comparison of the interferometric only (blue) and short-spacing corrected (orange) spectrum extracted along a line of sight in the central molecular region of M82. We find that the difference in integrated intensity is less than 3\% for individual sightlines, therefore, the SMA observations appear to recover the significant fraction of the total flux.  Integrated over the entire map, the missing integrated flux is still only around 10\%.

\section{Log-normal Model}\label{app: lognormal}

The gas density in a molecular cloud is described by 

\begin{equation}
    d\mathrm{P}\,(\ln n^{\prime})  \propto \mathrm{exp}\left(-\frac{(\ln n^{\prime} -\overline{\ln n^{\prime}})^{2}}{2\sigma^{2}}\right) d\,(\ln n^{\prime}),
\label{eq:pdf}
\end{equation} 
where $d\mathrm{P}$ is the fraction of molecules with volume densities in a logarithmic step $d\,\ln n^{\prime}$ centered on $n^{\prime}$, $n^{\prime} = n_{\mathrm{H}_{2}}/n_{0}$ is the volume density normalized by the mean volume density, $n_0$, and $\width$ is the width of the distribution \footnote{Note that the distribution does not peak at the center volume density ($n_{0}$), but rather it peaks at $\overline{\ln n^{\prime}}$.}. 

The adjusted intensities correspond to the sum of intensities weighted by the $\hdens$volume density PDF at fixed $\tkin$, $\coldens$, and isotopologue abundances as follows: 
\begin{equation}
I^{\mathrm{model}} = \frac{\int \hdens P\left(\hdens\right) I\left(\hdens, \coldens, \tkin, \xthirteen, \xeighteen \right)\,d\hdens}{\int \hdens P\left(\hdens \right) d\hdens}.
\label{eq:distributionintensity}
\end{equation}

\section{Matching Observations with Models}\label{app: likelihood}
For both models, to evaluate the goodness of fit for a parameter set (of structure $\theta~=~(\hdens, \tkin, \coldens, \xthirteen, \xeighteen, \sigma)$, where $\sigma$ is only considered in the log-normal model), we compute $\chi ^2$ as follows: \begin{equation}\label{eq:chi2}
    \chi^2(\theta) = \sum_{i=1}^n\left(\frac{\bff \cdot I_i^{\rm model} (\theta) -  c^{\rm line}_i\cdot I_i^{\rm obs}}{\sigma_i^{\rm obs}}\right)^2,
\end{equation}
where $c^{\rm line}_i$ considers the adjustment of the line width. We use $c^{\rm line}_i=10\,{\rm km\,s^{-1}}/{\rm FWHM_i}$, where the FWHM is the derived moment-2 value.  We adjust the modeled intensity by the beam filling factor of the line, $\bff$, which we vary from 0.01 -- 1 in logarithmic steps of 0.2\,dex for the one-zone model. We note that, in principle, one could also introduce, similarly, a \emph{global} beam filling factor for the log-normal density distribution. However, this would introduce an additional free parameter. Therefore, we refer the analysis of the introduction of a global beam filling factor to a future study once more CO isotopologue lines are available. The observational uncertainty is described by the parameter $\sigma_i$, not to be confused with log-normal distribution width $\sigma$. We additionally apply to the noise uncertainty a conservative estimate of 10\% uncertainty on the measured intensity, which is commonly adopted in the literature \citep[e.g.,][]{Leroy2017dens, Teng2022}. To quantify the significance of the minimum $\chi^2$, we compute for each parameter set a likelihood probability, assuming a multivariate Gaussian probability distribution:
\begin{equation}\label{eq:probdens}
    P(I^{\rm obs}|\theta) = \left(\prod_i^{n}(2\pi \sigma_i^2)^{-\frac{1}{2}} \right) \cdot e^{-\frac{1}{2}\chi^2(\theta)}
\end{equation}
We compute this probability for each parameter set in the model parameter space, which generates a 6D probability cube. We generate marginalized 1D and 2D likelihood distributions for any given parameter (or pair of parameters) by summing the joint probability distribution over the full range of parameter(s) except the one(s) in question. The resulting ``1DMax solution" parameter is then selected by determining the highest 1D likelihood in each parameter. For more details, see \cite{Teng2022}.

\begin{figure*}
    \centering
\includegraphics[width=0.8\textwidth]{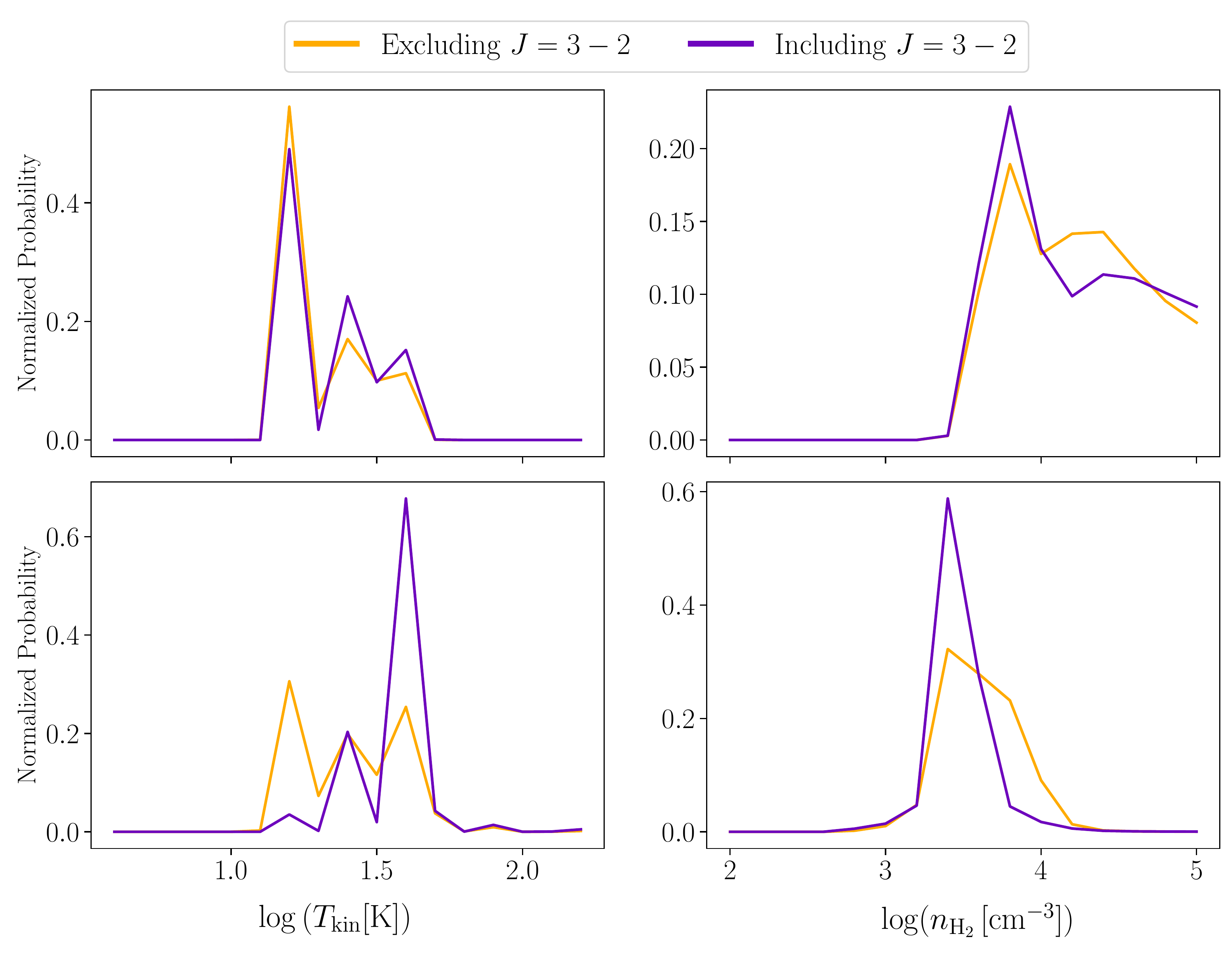}
    \caption{{\bf Overlayed 1D PDFs for $\tkin$ and $\hdens$.} We assess the effectiveness of the 3-2 line by overlaying 1D PDFs generated from likelihood analysis that was performed including (in purple) and excluding (in yellow) the 3-2 line. We present two examples, one (on the lower panel) in which the 3-2 line assists in tracing more extreme conditions, and another (the upper panel) in which the 3-2 line is not beneficial.}
    \label{fig:overlaypdfs}
\end{figure*}

\section{Examining effect of including higher J lines}\label{app: effectof32}

We examine how effectively CO isotopologue lines can trace more extreme molecular gas conditions, specifically, $\tkin$, a parameter that describes excitation conditions along with $\hdens$. This question has been examined by \cite{Teng2023}, who found that for three nearby barred galaxies, including the 3-2 line(s) from CO isotopologues resulted in much less scatter and/or bias when incorporated in the radiative transfer modeling.

We compare the one-zone model 1D PDFs derived from likelihood analysis including and excluding the $\co$ 3-2 line. In Figure \ref{fig:overlaypdfs}, we plot for two pixels the 1D PDFs for $\hdens$ and $\tkin$ parameter. These two specific pixels were chosen because they reflect the general behavior we see among all pixels. In purple, we plot the marginalization derived from including the $J = 3-2$ line, and in yellow, we plot the marginalization derived without the line. For the top panel, we see how, both parameters, including the higher $-J$ line do not change the resulting 1DMax solutions or the overall shape of the 1D PDF significantly. In other words, the 3-2 line does not help constrain the parameters further. 

For the bottom panel, we see how the $3-2$ line does not change the 1DMax solution for $\hdens$ (most likely due to the low critical density) but it does help trace higher temperatures. While this pixel demonstrates an example where the $3-2$ line aids in tracing more extreme excitation conditions, this behavior is not clear in all pixels. The $3-2$ line, for many pixels, is not effective in tracing the extreme star-forming conditions in the nucleus of M82.

\bibliography{references}{}
\bibliographystyle{aasjournal}
\end{document}